# Guidelines for measurement of thermal emission

Yuzhe Xiao[1], Chenghao Wan[1,2], Alireza Shahsafi[1], Jad Salman[1], Zhaoning Yu[3,1], Raymond Wambold[1], Hongyan Mei[1], Bryan E. Rubio Perez[1], Chunhui Yao[1], and Mikhail A. Kats[1,2,3*]

[1]*Department of Electrical and Computer Engineering, University of Wisconsin-Madison, Madison, Wisconsin 53706, USA*
[2] *Department of Materials Science and Engineering, University of Wisconsin-Madison, Madison, Wisconsin 53706, USA*
[3]*Department of Physics, University of Wisconsin-Madison, Madison, Wisconsin, USA 53706*

*Email address: mkats@wisc.edu*

Thermal emission is the radiation of electromagnetic waves from hot objects. The promise of thermal-emission engineering for applications in energy harvesting, radiative cooling, and thermal camouflage has recently led to renewed research interest in this topic. There is a substantial need for accurate and precise measurement of thermal emission in a laboratory setting, which can be challenging in part due to the presence of background emission from the surrounding environment and the measurement instrument itself. This is especially true for measurements of emitters at temperatures close to that of the environment, where the impact of background emission is relatively large. In this paper, we describe, recommend, and demonstrate general procedures for thermal-emission measurements that are applicable to most experimental conditions, including less-common and more-challenging cases that include thermal emitters with temperature-dependent emissivity and emitters that are not in thermal equilibrium.

## I: Introduction

Thermal emission is a fundamental phenomenon for which any object with temperature greater than absolute zero emits energy in the form of electromagnetic waves. The thermal emission from a blackbody (an ideal thermal emitter) only depends on its temperature, as described by Planck's law [1]. For a non-ideal thermal emitter, *i.e.*, a gray body, thermal emission also depends on how effectively the emitter can radiate, quantified as its emissivity. For an object in thermal equilibrium, the emissivity is exactly equal to its absorptivity, as described by Kirchhoff's law [2]. Therefore, engineering of thermal emission is equivalent to engineering the optical absorption, which can be accomplished, *e.g.*, using the toolkits of nanophotonics and materials engineering [3]. Engineered thermal emission has many applications, including energy harvesting [4], lighting [5], radiative cooling [6], and thermal camouflage [7].

One important aspect of this research field is the experimental characterization of thermal emitters. For an emitter in thermal equilibrium where Kirchhoff's law applies, its emissivity can be indirectly obtained by measuring the absorption, *e.g.*, via a combination of reflection, transmission, and scattering measurements [4],[5]. For non-scattering emitters, this approach is often more favorable than direct thermal-emission measurement because measurement of reflectance and transmittance is well established. However, there are many situations where such indirect measurements are not the best choice. For example, measurement of the actual radiated light may be more favorable for thermal emitters that are scattering [10] or highly absorbing (*i.e.,* the reflection and transmission are both small) [11]. This may also be the case when a focusing system is used (*e.g.,* to measure small samples). Thermal-emission measurements are often more robust against focusing errors compared to reflection and transmission measurements because the measured thermal emission signal is proportional to the product of area and solid angle (*i.e.*, etendue [12]), which is a conserved quantity in an imaging system. Furthermore, there are situations where Kirchhoff's law does not apply (at least not in its conventional form), and direct thermal-emission measurement is the only way to characterize the emitters. These situations include emitters that are not in thermal equilibrium [13]–[15] or are not reciprocal [16], [17]. In practice, even in cases where Kirchhoff's law does apply, it is often beneficial to perform both indirect and direct measurement for maximum confidence in the measurement results[1].

Directly measuring thermal emission from a sample can be challenging in part because almost every component of the instrument and the surrounding environment thermally radiates light, resulting in a background signal that may be difficult to isolate. This is especially true when the emitter temperature is not significantly higher than that of the ambient environment [18]. Depending on the properties of the thermal emitters and the experimental conditions, different methods have been proposed and used to characterize the emissivity [19]–[26]. However, to the best of our knowledge, there is no comprehensive summary of thermal-emission-measurement methodology in the existing literature. In this paper, we provide general guidelines for thermal-emission measurements in a variety of experimental conditions. For example, we describe how to directly measure low-temperature emitters that have temperature-dependent emissivity, where conventional methods cannot be easily applied. These guidelines should apply to thermal-emission measurement systems that use spectrometers based on dispersive elements such as gratings and prisms, and spectrometers based on interference, such as Fourier-transform spectrometers (FTSs or FTIRs), though we only provide experimental examples based on the FTS in our laboratory.

---

[1] As an example, in our setup with an infrared microscope, we use a liquid-nitrogen-cooled photoconducting detector that is substantially nonlinear, which complicates our temperature-dependent reflection measurements. In such measurements, the thermal-emission power from high-temperature samples leads to a partial saturation of the detector, which decreases the detector response and results in an observed (nonphysical) decreased value of reflectance at high temperatures.

## II: Overview

Figure 1(a) is a simplified schematic of a typical thermal-emission measurement setup. The thermal emitter to be characterized is heated to a certain temperature with a heat stage, and the corresponding thermal emission (red arrows) is measured using a spectrometer. The spectrometer may be based on dispersive elements, such as gratings and prisms, but the majority of thermal-emission measurements are performed in the mid infrared using Fourier-transform spectrometers (FTSs) [19]–[26]. FTSs are often preferred for mid-infrared measurements due in part to their "multiplex advantage" over dispersive instruments [27].

In Figure 1(b), we show a schematic of the FTS in our laboratory—which we will use throughout this paper to demonstrate the various measurement approaches. Thermal emission from the sample (red arrows) is collected with a parabolic mirror (numerical aperture ~0.05) and then passes through a Michelson interferometer with a moving mirror, and the signal is modulated by the interferometer and recorded using a liquid-nitrogen-cooled mercury-cadmium-telluride (MCT) detector. The signal from the detector (*i.e.,* the interferogram) is Fourier transformed to obtain the emission signal in the frequency domain [28]. In our multi-purpose setup, the beam path between the interferometer and detector includes several mirrors and apertures ["Optical components" in Fig. 1(b)].

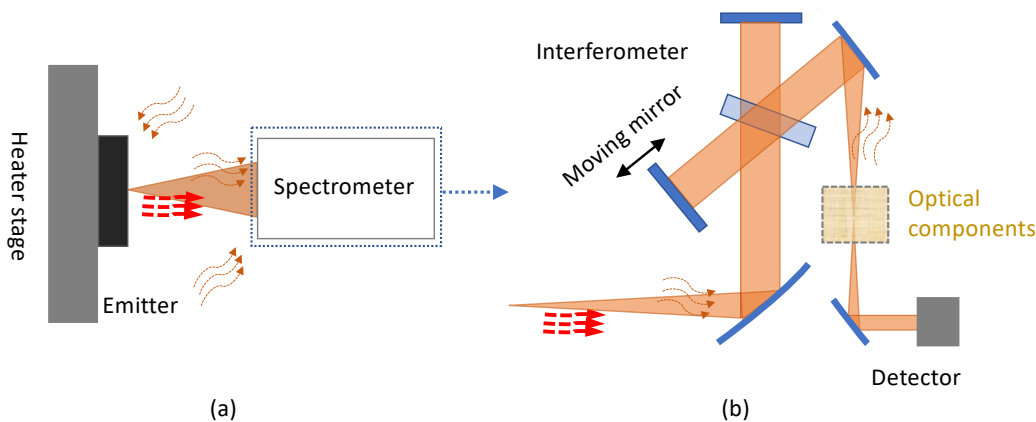

**Figure 1**. (a) Simplified schematic of a typical thermal-emission measurement setup. (b) Schematic of our Fourier-transform spectrometer (FTS) [modified Bruker VERTEX 70]. The samples were affixed to a temperature-controlled stage mounted inside the FTS sample compartment and rotated by 10° with respect to the beam path to avoid multiple reflections between the samples and the interferometer. In both (a) and (b), red arrows represent the emission from the sample, while the brown arrows represent background emission.

Unlike the measurement of other types of emission, such as fluorescence where the signal is relatively narrowband and the background can be easily separated (*e.g.,* visible fluorescence measured in a dark lab room while spectrally filtering out the pump wavelength), thermal-emission measurements are often complicated by the presence of a thermal background that overlaps both spectrally and spatially with the sample emission (brown arrows in Fig. 1). The background can come from both the surrounding environment and the instrument itself; in the case of an FTS, the background can originate from components both before and after the interferometer [Fig. 1(b)]. More importantly, the background emission contribution to the measured signal can be sample dependent, *i.e.*, various components of the background can be reflected or scattered by the emitter into the measurement beam path. Finally, the background can also be temperature dependent, if the optical properties of the emitter change with temperature and the background emission somehow interacts the sample (*e.g.*, via reflection or scattering) before making its way to the detector. Therefore, in general, the measured signal, $S_x(\lambda, T)$, from an emitter $x$ at temperature $T$, can be expressed as:

$$S_x(\lambda, T) = m(\lambda)[I_x(\lambda, T) + B_x(\lambda, T)], \tag{1}$$

where $m(\lambda)$ is the wavelength-dependent system-response function that quantifies the collection efficiency of the beam path and the detector response, and $I_x(\lambda, T)$ and $B_x(\lambda, T)$ are the true emission signal and background for emitter $x$, respectively. Note that there may be some background-emission sources located within the interferometer, for which the system response is not necessarily $m(\lambda)$; we group this term into $B_x(\lambda, T)$ without loss of generality.

In principle, it is desirable to use cooled detectors (*e.g.*, MCT detectors at liquid nitrogen temperatures) due to their much higher sensitivity compared to uncooled detectors. Nevertheless, room-temperature detectors [*e.g.*, those based on doped triglycine sulfate (DTGS)] have one surprising advantage over cooled detectors for thermal-emission measurements. If the source of the background emission is at room temperature, the background will be in thermal equilibrium with the detector, and will thus not be detected [*i.e.*, the measured $B_x(\lambda, T) = 0$], greatly simplifying the data analysis. Indeed, a number of thermal-emission experiments have been performed with room-temperature detectors [7], [26]. This approach, however, requires the detector temperature to be precisely the same as the temperature of all the source of the background (which may not be possible if not all background sources are at precisely the same temperature), and also means that the signal-to-nose ratio will likely be lower compared to a setup with cooled detectors [29]. Thus, our recommendation is to use room-temperature detectors when the signal-to-noise ratio is sufficient and the results need not be completely accurate. For high-accuracy or low-power (*i.e.*, low temperature or small area) measurements, a cooled detector should be used. The focus of this paper is on thermal emission measurements using cooled detectors, though the descriptions will also be valid for uncooled detectors.

If the emitter is in thermal equilibrium (*i.e.,* the emitter has a single uniform temperature $T$), then the true emission signal can be written as: $I_x(\lambda, T) = \epsilon_x(\lambda, T)I_{BB}(\lambda, T)$, where $\epsilon_x(\lambda, T)$ is the emissivity for emitter $x$ at $T$ (in kelvin) and $I_{BB}(\lambda, T)$ is the blackbody-radiation distribution given by Planck's law:

$$I_{BB}(\lambda, T) = \frac{2hc^2}{\lambda^5} \frac{1}{e^{\frac{hc}{\lambda k_B T}} - 1}. \tag{2}$$

Here $h$, $c$ and $k_B$ are the Planck constant, speed of light in vacuum, and the Boltzmann constant, respectively. Therefore, in the case of thermal equilibrium, Eq. 1 becomes

$$S_x(\lambda, T) = m(\lambda)[\epsilon_x(\lambda, T) I_{BB}(\lambda, T) + B_x(\lambda, T)]. \tag{3}$$

Since most of the thermal emitters in the research literature are at thermal equilibrium, the focus of a typical thermal-emission measurement is to find $\epsilon_x(\lambda, T)$ from the measured signal $S_x(\lambda, T)$.

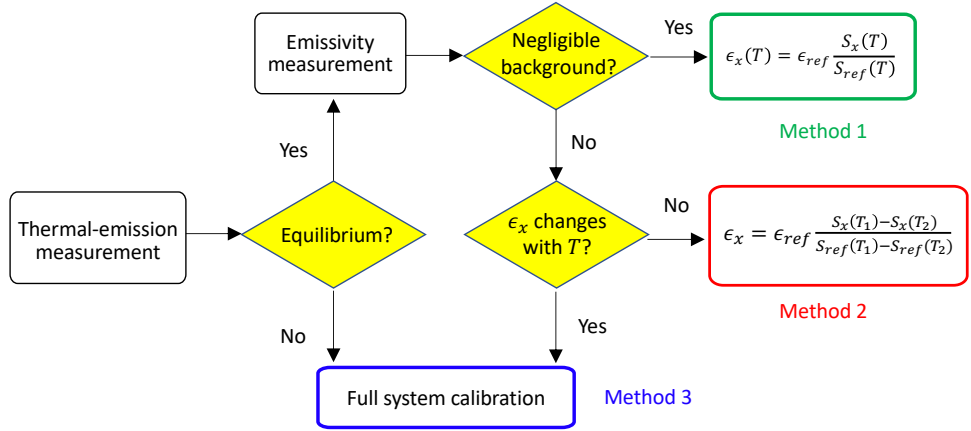

**Figure 2**. Flowchart for direct thermal-emission measurements. If the emitter is in thermal equilibrium (*i.e.*, can be described with a single temperature, $T$), a thermal-emission measurement is equivalent to measurement of the emissivity. Depending on the relative magnitude of the background and whether the emissivity changes with temperature, different branches of the flowchart should be followed. If the emitter is not in thermal equilibrium, then a full system calibration is necessary. Note that for cases where Kirchhoff's law applies (thermal equilibrium, top part of the flowchart), emissivity can also be obtained via indirect measurements.

Figure 2 provides a flowchart that can be used as a guide for emission measurements. If the emitter is in thermal equilibrium, measuring thermal emission $I_x(\lambda, T)$ provides the same information as measuring emissivity $\epsilon_x(\lambda, T)$ because they are related by Planck's law. Measuring emissivity is often easier than measuring the thermal-emission spectrum, even without the use of Kirchhoff's law. Depending on the relative magnitude of background emission and whether the emissivity changes with temperature, different methods should be used to measure the emissivity.

If the background emission can be neglected, the emissivity can be obtained by taking the ratio of measured emission signal from the sample emitter, $S(\lambda, T)$, to that from a known reference, *i.e.*, $\epsilon_x(\lambda, T) = \epsilon_{ref}(\lambda, T) \frac{S_x(\lambda,T)}{S_{ref}(\lambda,T)}$. We denote this as *Method 1*, which is used when the emitter temperature is much higher than the surrounding environment, for example in the case of emitters in thermal photovoltaic systems [30]– [32], or if the measurement system itself has very low background thermal emission (or when the background is canceled out because the detector is at the same temperature as the background sources). We discuss this case in Sec. III, where we also provide a simple procedure to estimate the relative magnitude of the background.

If the background emission is not negligible but the emissivity does not change with temperature, the emissivity of the sample can be obtained by measuring the thermal emission from both the sample and the reference at two different temperatures $(T_1, T_2)$ via $\epsilon_x(\lambda) = \epsilon_{ref}(\lambda)\frac{S_x(\lambda, T_1) - S_x(\lambda, T_2)}{S_{ref}(\lambda, T_1) - S_{ref}(\lambda, T_2)}$. We denote this as *Method 2*, which is widely used in remote sensing where the thermal-emission signal is not significantly higher than the background [19]–[21]. We discuss this case in Sec. IV, where we also show that *Method 2* cannot be applied to the case of temperature-dependent emissivity.

For the scenario where the background emission cannot be neglected and the emissivity of the sample is temperature dependent, a full calibration of the measurement setup, quantifying the system response, $m(\lambda)$, and the background, $B_x(\lambda, T)$, is needed. We refer this as *Method 3*, which has been used to obtain the emissivity for opaque and non-scattering emitters when the emissivity is temperature-independent [33]–[35]. We further develop *Method 3* for semitransparent and scattering emitters in this work. Example uses of *Method 3* include measuring thermal emitters that are based on phase-change materials such as vanadium dioxide ($VO_2$) [36] and germanium-antimony-tellurium (GST) [37]. As an example, in Sec. V, we apply *Method 3* to measure the emissivity of a $VO_2$-based emitter.

Finally, for emitters that are not in thermal equilibrium, the concept of emissivity is not applicable, and one must measure the total emission using *Method 3*. In Sec. V, we demonstrated such a non-equilibrium thermal-emission measurement on a fused-silica slab with a gradient temperature distribution.

### III: Emissivity measurements with negligible background

When the background is negligible compared to the sample emission [*i.e.*, $B_x(\lambda, T) \ll I_x(\lambda, T)$ in Eq. 1], the measurement of emissivity is relatively straightforward. This is the case if the temperature of sample is very high [*i.e.*, $I_x(\lambda, T)$ is large] or if the measurement system itself has very low background [*i.e.*, $B_x(\lambda, T)$ is small]. In this case, Eq. 3 can be well approximated by

$$S_x(\lambda, T) = m(\lambda)\epsilon_x(\lambda, T)I_{BB}(\lambda, T). \tag{4}$$

Therefore, $\epsilon_x(\lambda, T)$ can be obtained simply by measuring the thermal-emission signal for both the sample $x$ and a known reference $\alpha$, and taking the ratio:

$$\epsilon_x(\lambda, T) = \epsilon_\alpha(\lambda, T)\frac{S_x(\lambda, T)}{S_\alpha(\lambda, T)}. \tag{5}$$

Equation 5, which we denote as *Method 1*, is widely used by the nanophotonics community [22]–[26] because a significant portion of the thermal emitters of interest are for thermophotovoltaics applications and are thus characterized at high temperatures. In principle, the reference $\alpha$ can be any emitter if its thermal emissivity at temperature $T$ is known. In practice, reference samples with sharp spectral features or low emissivity can introduce artifacts in the extracted sample emissivity; therefore, laboratory blackbodies, which have a temperature- and wavelength-independent emissivity close to unity, are usually chosen for convenience. Typical laboratory blackbodies include enclosures with absorbing walls and a small hole through which light can enter and exit [12] and highly absorbing optical structures that suppress reflection and scattering, such as vertically oriented carbon nanotube (CNT) arrays [38] or black soot [26].

Emissivity characterization via *Method 1* is very simple, but it is important to make sure that the background is indeed much smaller than the sample emission; otherwise, the measured emissivity will be incorrect. Here, we provide a simple procedure to estimate the relative magnitude of the background. To do so, we can calculate the ratio of the measured thermal-emission spectra from a sample $\alpha$, $S_\alpha(\lambda, T)$, at two temperatures, $T_1$ and $T_2$:

$$\rho_\alpha(\lambda, T_1, T_2) = \frac{S_\alpha(\lambda, T_1)}{S_\alpha(\lambda, T_2)} = \frac{\epsilon_\alpha(\lambda, T_1)I_{BB}(\lambda, T_1) + B_\alpha(\lambda, T_1)}{\epsilon_\alpha(\lambda, T_2)I_{BB}(\lambda, T_2) + B_\alpha(\lambda, T_2)} \qquad (6)$$

For $B_\alpha(\lambda, T_{1,2}) \ll \epsilon_\alpha(\lambda, T_{1,2})I_{BB}(\lambda, T_{1,2})$ and $\epsilon_\alpha(\lambda, T_1) \approx \epsilon_\alpha(\lambda, T_2)$, a theoretical ratio can be obtained from Planck's law:

$$\rho_{Planck}(\lambda, T_1, T_2) = \frac{I_{BB}(\lambda, T_1)}{I_{BB}(\lambda, T_2)} \qquad (7)$$

The relative magnitude of the background can be estimated by comparing $\rho_\alpha$ in Eq. 6 from the measurements with $\rho_{Planck}$ in Eq. 7, which are approximately equal only if the background can be neglected and the sample emissivity does not change with temperature. If $\rho_\alpha(\lambda, T_1, T_2) \neq \rho_{Planck}(\lambda, T_1, T_2)$, then it can be deduced that either the background cannot be neglected or the sample has temperature-dependent emissivity.

For samples with constant emissivity, $\rho_\alpha(\lambda, T_1, T_2) \neq \rho_{Planck}(\lambda, T_1, T_2)$ implies a non-negligible background. In addition, the sign of the difference of the experimental ratio $\rho_\alpha$ and the theoretical ratio $\rho_{Planck}$ yields the sign of the background, which may not be positive [18][39].

Intuitively, one would expect background to always be positive. Indeed, this should always be the case for measurement systems based on dispersive elements, such as gratings or prisms. However, this is not always the case in thermal-emission measurements with an FTS [18][39]. In FTSs, the spectra are obtained by Fourier-transforming the interferogram [28], where the interferogram from the background originating from components after the interferometer can have a $\pi$ phase difference with respect to the interferogram that encodes the emission from the sample. This $\pi$ phase difference of the interferogram appears as a negative background in the Fourier-transformed spectrum [18].

To demonstrate the procedure to estimate the background, we performed measurements of thermal emission from a polished sapphire wafer using our FTS setup shown in Fig. 1(b) and plotted the measured signal at 373, 423, 523, and 573 K in Fig. 3(a). The ratios of the measured emission signal (Eq. 6) together with the theoretical ratios from Planck's law corresponding to the same temperature differences (Eq. 7) are plotted in Fig. 3(b). Except for some wavelength regions near 16 μm, the experimental ratios are higher than the theoretical ones. Thus, because $T_1 > T_2$, $B(\lambda, T)$ must be negative in our setup for this sample. The thermal background is quite large in our particular system, which is convenient for the demonstration of the calibration methods in this paper, though somewhat inconvenient for thermal-emission measurements in general.

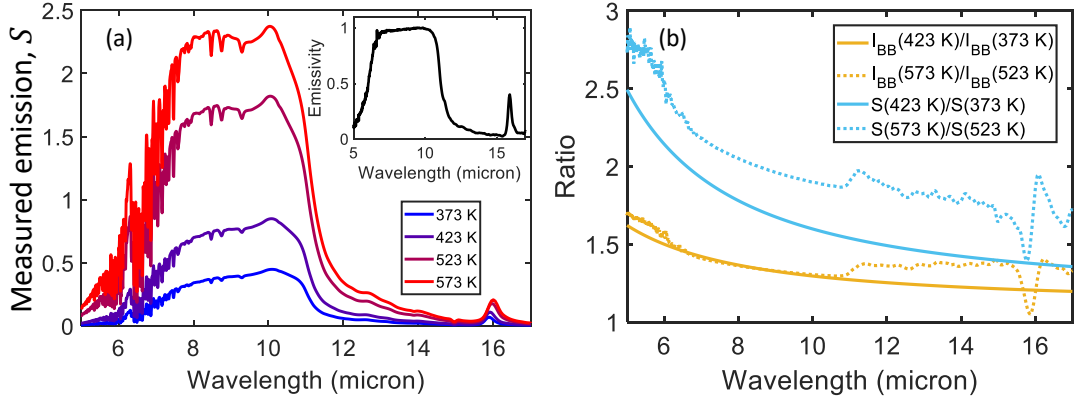

**Figure 3**. (a) Measured thermal-emission signal (in arbitrary units) from a polished sapphire wafer at four different temperatures. The measurements were unpolarized, at an angle of 10°. The sharp features in the spectra (especially between 5 and 8 μm) are due to absorption of the ambient air. In the inset, we plotted the measured emissivity of this sapphire wafer. (b) Ratios of the experimentally measured signal are shown using dotted curves (red: $S_{573\,K}/S_{523\,K}$ and blue: $S_{423\,K}/S_{373\,K}$). Solid curves show the ratios from Planck's law at the same temperatures.

The magnitude of the difference between the experimental and theoretical ratios is smaller at high sample temperatures [$S_{573\,K}/S_{523\,K}$, Fig. 3(b)] than that at low temperatures [$S_{423\,K}/S_{373\,K}$, Fig. 3(b)], which is expected since the background does not change with $T$ while the sample emission increases with $T$. The difference also changes with wavelength. Specifically, for the high-temperature ratios in Fig. 3(b), we find the experiment (dotted) quite closely matches the theory (solid) in the spectral range of 6 to 11 μm, where the sapphire emissivity is very close to one [Fig. 3(a), inset]. A considerable difference shows up outside of this spectral range. This is expected because the condition $B_\alpha \ll \epsilon_\alpha I_{BB}$ is only met at these temperatures in the spectral region where $\epsilon_\alpha$ is sufficiently large.

# IV: Emissivity measurements with non-negligible background

There are several categories of thermal-emission measurements that are likely to have a substantial background that cannot be neglected, including remote sensing of temperature [19]–[21] and characterization of engineered low-temperature thermal emitters [40]–[42].

**(1): Temperature-independent emissivity**

Most thermal emitters have optical properties that do not change very much with temperature over a broad temperature range. Making the assumption of a temperature-independent emissivity, Eq. 3 becomes

$$S_x(\lambda, T) = m(\lambda)\{\epsilon_x(\lambda)I_{BB}(\lambda, T) + B_x(\lambda)\}. \tag{8}$$

Therefore, it is possible to cancel out the background $B_x(\lambda)$ by subtracting the measured signal of an emitter $x$ at two temperatures $T_1$ and $T_2$:

$$S_x(\lambda, T_1) - S_x(\lambda, T_2) = m(\lambda)\epsilon_x(\lambda)\{I_{BB}(\lambda, T_1) - I_{BB}(\lambda, T_2)\}. \tag{9}$$

Note that Eq. 9 has a similar form to Eq. 4: the difference of the measurements at two temperatures is proportional to the sample emissivity. Thus, using a known reference $\alpha$ with temperature-independent emissivity, we can obtain $\epsilon_x(\lambda)$ by:

$$\epsilon_x(\lambda) = \epsilon_\alpha(\lambda) \frac{S_x(\lambda, T_1) - S_x(\lambda, T_2)}{S_\alpha(\lambda, T_1) - S_\alpha(\lambda, T_2)}. \tag{10}$$

We refer to Eq. 10 as *Method 2*. This is a commonly used method for the measurement of emissivity, especially in the remote sensing community [19]–[21], where remote emission measurements can be used to sense, *e.g.*, temperature and atmospheric humidity. Several slightly different methods have been used for calibrating thermal emissivity [43]–[45], but all of these are essentially the same as *Method 2*: the background is removed by subtracting thermal emission at two temperatures.

Using *Method 2*, the measurement noise can become an issue, depending on the difference between $T_1$ and $T_2$. If the emissivity does not change with temperature at all, then $T_1$ and $T_2$ should be chosen far away from each other so that the denominator in Eq. 10 is sufficiently large. If, on the other hand, the emissivity only remains constant within a narrow temperature window, then $T_1$ and $T_2$ must be chosen accordingly. In this case, the difference between temperatures will be small, which can amplify the measurement noise due to the small value of the denominator in Eq. 10 and may yield unreliable results [*e.g.*, Fig. 5(d) and (e) for small temperature differences]. Note that *Method 2* will break down if the sample emissivity changes considerably with temperature even we pick $T_1$ and $T_2$ close to each other; for more discussion, see Sec. IV(2).

To validate *Method 2* using our setup, we characterized the emissivity of polished sapphire and fused-silica wafers (Fig. 4). Here, a laboratory blackbody (approximately 500-μm tall vertically aligned carbon nanotube (CNT) forest [38] grown on a silicon substrate) with a constant emissivity of ~0.97 [18] was used as the reference. We plot the measured thermal-emission signal from the CNT blackbody, sapphire, and fused silica at 323 and 348 K in Fig. 4(a). For the liquid-nitrogen-cooled MCT detector in our FTS, a temperature difference of 25 K was sufficient.

Using the spectra in Fig. 4(a) and Eq. 10, we calculated the emissivity of sapphire and fused silica, and plotted them in Fig. 4(b). As shown in Fig. 4(b), the emissivity of the fused-silica wafer is very close to unity except for the spectral region near 9 μm. The dip in the emissivity is related to the vibrational resonances of silica [46]. The emissivity of sapphire is very close to unity from 7 to 11 μm and rapidly decreases outside of this spectral region. The emissivity peak near 16 μm is related to the vibrational resonances of sapphire [47]. Note that the emissivity of the CNT "blackbody" is actually slightly smaller than the emissivities of the wafers in some spectral regions (*e.g.,* 6–8 μm for fused silica and 8–10 μm for sapphire).

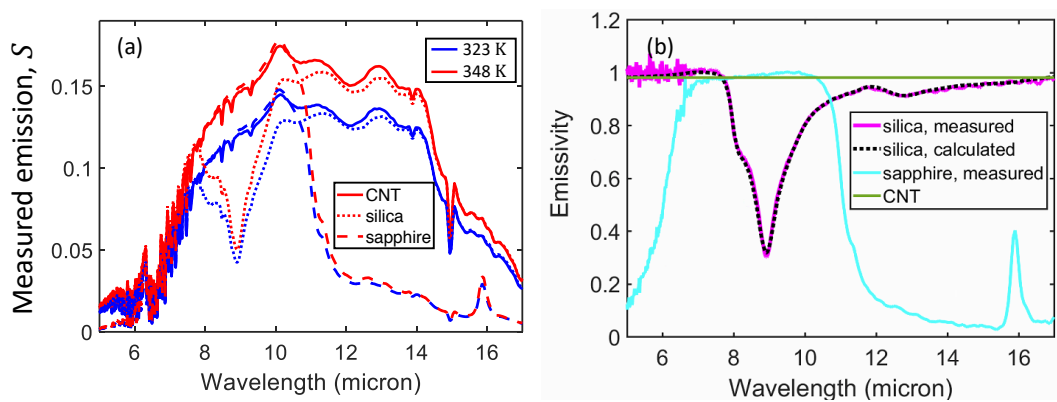

**Figure 4**. (a) Measured thermal-emission signal (in arbitrary units) from a CNT blackbody (solid), a fused-silica wafer (dotted), and a sapphire wafer (dashed), at 323 (blue) and 348 K (red). The measurements were unpolarized, at an angle of 10°. (b) Measured emissivity of the fused-silica (solid purple) and sapphire (solid cyan) wafers using the data in (a), extracted via Eq. 10 (*Method 2*). The emissivity of the CNT blackbody is plotted using the solid-green curve. The calculated emissivity of fused silica using $n$ and $\kappa$ measured with variable-angle spectroscopic ellipsometry (VASE) is plotted using the dotted-black curve, showing excellent agreement with *Method 2*.

To confirm the results of *Method 2*, we also calculated the emissivity of the fused-silica wafer indirectly using Kirchhoff's law. Because the fused-silica wafer is opaque for $\lambda > 5$ μm and our sample is non scattering (the wafer is polished), Kirchhoff's law yields:

$$\epsilon(\lambda) = A(\lambda) = 1 - R(\lambda), \tag{11}$$

where $A(\lambda)$ is the sample absorptivity. We measured the optical properties ($n$ and $\kappa$) of the fused-silica wafer using variable-angle spectroscopic ellipsometry (VASE) and then calculated its reflectance using Fresnel coefficients. The indirectly calculated emissivity via Eq. 11 is plotted in Fig. 4(b) and is in excellent agreement with *Method 2*.

## (2): Breakdown of *Method 2* given temperature-dependent emissivity

Somewhat counterintuitively, *Method 2* (Eq. 10) fails for emitters with temperature-dependent emissivity (*e.g.*, emitters based on phase-change materials [7], [26], [48]–[50]), if the rate of emissivity change with temperature is larger than a certain value that is related to rate of change of the Planck distribution with temperature. We provide the mathematical proof in Appendix I and show this breakdown experimentally as follows.

We used a thermal emitter comprised of a ~100 nm film of vanadium dioxide (VO$_2$) sputtered on a sapphire substrate (more details about this sample can be found in Ref. [48]). This VO$_2$-based emitter provides an excellent test of *Method 2*, because we can select one temperature range where the emissivity changes rapidly (around 340 K) and a different temperature range where the emissivity is almost independent of temperature (around 370 K).

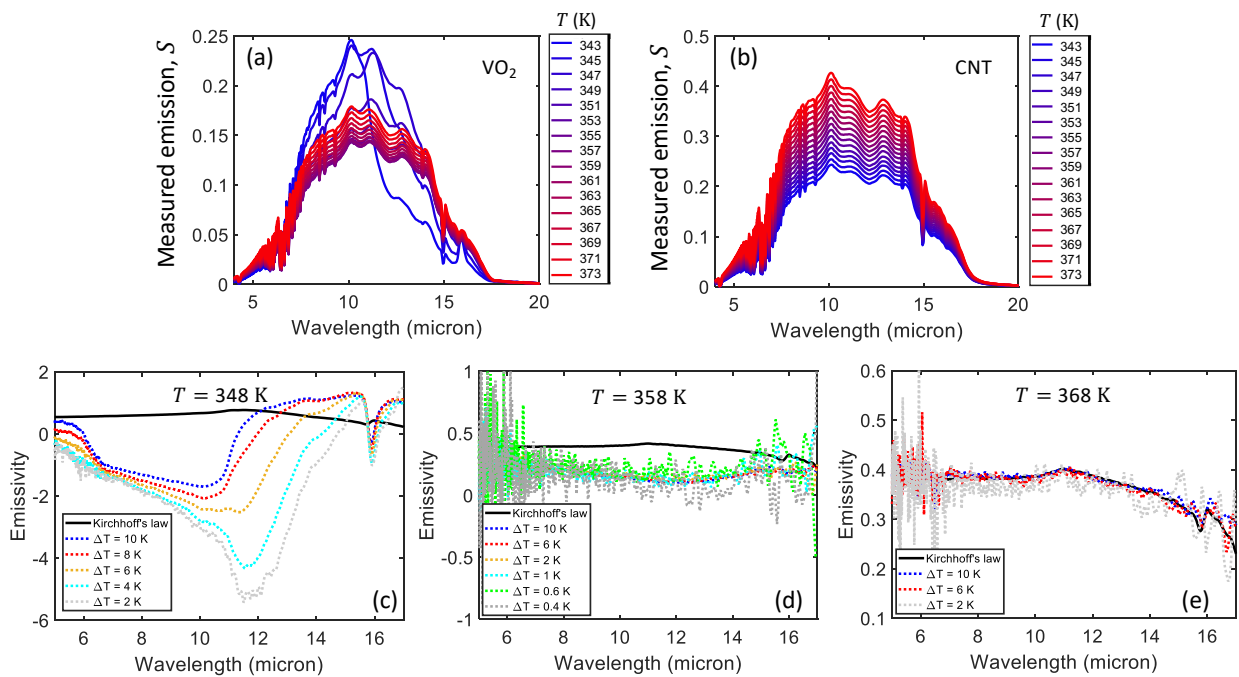

**Figure 5**. (a-b) Measured thermal-emission signal (in arbitrary units) from the VO$_2$-based emitter and the blackbody from 343 to 373 K. The measurements were unpolarized, at an angle of 10°. (c-e) Comparison of the measured emissivity of the VO$_2$-based emitter using Kirchhoff's law (solid black, here assumed to be correct) and direct emission measurements analyzed with *Method 2* (dotted curves, Eq. 12) around $T = 348$ (c), 358 (d), and 368 K (e) for different

values of $\Delta T$. The results in (c) and (d) show that *Method 2* cannot be applied to emitters with temperature-dependent emissivity. The agreement in (e) comes from the fact that the emissivity of the VO$_2$-based emitter no longer changes with temperature beyond 360 K.

Figure 5(a) shows the measured thermal-emission signal from the VO$_2$-based emitter for temperatures from 343 to 373 K. We used a laboratory CNT blackbody as the reference and plotted its measured thermal-emission signal at the same temperatures in Fig. 5(b). For the VO$_2$-based emitter, the emission signal first decreases rapidly from 343 to 353 K and then increases gradually as the temperature is further increased, due to the change in emissivity near the insulator-to-metal transition of VO$_2$ [31] [48] (also see Sec. V, Fig. 8 below). When $T > 363$ K, the emissivity hardly changes with temperature and the thermal-emission spectrum increases with temperature at a roughly constant rate, similar to the blackbody.

We used *Method 2* (Eq. 10) to attempt to obtain the emissivity of the VO$_2$-based emitter, selecting $T_1$ and $T_2$ with different $\Delta T$, i.e., $T_1 = T + \Delta T/2, T_2 = T - \Delta T/2$:

$$\epsilon_x(\lambda, T) \stackrel{?}{=} \epsilon_\alpha(\lambda) \frac{S_x(\lambda, T+\Delta T/2) - S_x(\lambda, T-\Delta T/2)}{S_\alpha(\lambda, T+\Delta T/2) - S_\alpha(\lambda, T-\Delta T/2)}. \tag{12}$$

We put a question mark in Eq. 12 because, as described in Appendix I, we anticipate that it breaks down for emitters with temperature-dependent emissivity. To verify this breakdown, we tested Eq. 12 at three central temperatures: $T = 348, 358$ and $368$ K, using different values of $\Delta T$, and plotted the results in Figs. 5(c)-(e). To validate the results, we also measured the emissivity indirectly at these temperatures via Kirchhoff's law (Eq. 11) by heating the sample. Note that Eq. 11 can be used here because the VO$_2$-based emitter is opaque and non-scattering in this wavelength range [48].

As shown in Fig. 5(c), the measured emissivity deviates substantially from the Kirchhoff's-law measurement and is clearly unphysical. These unphysical results occur because the emissivity of the VO$_2$-based emitter changes rapidly with temperature around 340 K [48]. Conversely, the measured emissivity at 368 K [Fig. 5(e)] overlaps quite well with the result from Kirchhoff's law measurement, even for a relatively large $\Delta T = 10$ K. The emissivity of the VO$_2$-based emitter hardly changes for $T > 360$ K [48], and therefore Eq. 12 returns correct results. Also, note that as $\Delta T$ decreases below 2 K, the noise in emissivity becomes quite significant. Figure 5(d) shows the results in the intermediate case of $T = 358$ K. Near this temperature, the emissivity changes with temperature, but relatively slowly [48]. The results from Eq. 12 deviate from the Kirchhoff's law measurement, even for a very small $\Delta T$ of 0.4 K.

The results in Fig. 5(c-e) can be explained via the following equation (Eq. A9 in Appendix I), which quantifies the expected difference between the extracted and true emissivity (*i.e.*, the expected error) using *Method 2* (Eq. 12) with our FTS, in the limiting case of $\Delta T \to 0$ K:

$$\Delta\epsilon_x(\lambda, T) = H(\lambda, T)\frac{\partial \epsilon_x(\lambda,T)}{\partial T}, \qquad (13)$$

where $H(\lambda, T)$ is a positive function that remains roughly constant for the temperature range discussed here (Fig. A1 in Appendix I). In the temperature range where the emissivity of the $VO_2$-based emitter changes very fast with temperature (*i.e.*, 343 to 353 K), a large error is observed in Fig. 5(c), while a relative smaller error is observed in Fig. 5(d) when the emissivity changes slower with temperature (*i.e.*, $T > 353$ K). When $T > 363$ K, the change in emissivity is very small, leading to very small errors in Fig. 5(e).

We emphasize that if the background is negligible, *Method 1* can still be used even for temperature-dependent emissivity. However, in the case of non-negligible background, a different approach must be taken. In the following section, we describe one such approach.

## V: Full calibration of the measurement setup

In this section, we present a method that is more complex than *Methods 1* and *2* but can handle most thermal-emission measurements, which is especially useful for characterizing emitters whose emissivity changes with temperature and emitters that are not in thermal equilibrium. This method allows for the extraction of the true emission $I_x(\lambda, T)$ by first fully characterizing the system response $m(\lambda)$ and the background $B_x(\lambda, T)$:

$$I_x(\lambda, T) = \frac{S_x(\lambda,T)}{m(\lambda)} - B_x(\lambda, T). \qquad (14)$$

If the emitter is in thermal equilibrium, its emissivity can be determined via:

$$\epsilon_x(\lambda, T) = \frac{\frac{S_x(\lambda,T)}{m(\lambda)} - B_x(\lambda,T)}{I_{BB}(\lambda, T)}. \qquad (15)$$

The response function $m(\lambda)$ can be obtained from Eq. 9 by using the thermal-emission signal measured from a known reference $\alpha$ whose optical properties do not change with temperature at two different temperatures $T_1$ and $T_2$:

$$m(\lambda) = \frac{S_\alpha(\lambda, T_1) - S_\alpha(\lambda, T_2)}{\epsilon_\alpha(\lambda)[I_{BB}(\lambda, T_1) - I_{BB}(\lambda, T_2)]} \qquad (16)$$

The characterization of $B_x(\lambda, T)$ is more complicated. Depending on both the optical properties of the emitter and the specific experimental setup, there are several ways that the emitter $x$ could have an impact on $B_x(\lambda, T)$, such as via reflection, transmission, or scattering. For example, some background emission $b(\lambda)$ can be transmitted through non-opaque emitters and the corresponding portion of the detected background would be $Tr_x(\lambda, T)b(\lambda)$, where $Tr_x(\lambda, T)$ is the transmittance of the emitter. We first discuss the most common and simplest case of opaque and non-scattering emitters.

**(1) Characterization of opaque and non-scattering emitters**

For an opaque and non-scattering emitter, the sample-dependence of background is expected to be entirely due to light reflected by the sample into the beam path. Therefore, $B_x(\lambda, T)$ can be written as:

$$B_x(\lambda, T) = B_1(\lambda) + R_x(\lambda, T)B_2(\lambda), \tag{17}$$

where $R_x(\lambda, T)$ is the reflectance of the emitter and $R_x(\lambda, T)B_2(\lambda)$ and $B_1(\lambda)$ represent the sample-dependent and sample-independent components of the background, respectively. Note that here we consider tilted samples that do not directly face the beam path; for non-scattering samples directly facing the beam path there may be no reflected background, but the system becomes more complicated due to the potential for multiple reflections between the instrument and the sample.

A full characterization of the system therefore involves finding both $B_1(\lambda)$ and $B_2(\lambda)$. To do this, at least two temperature-independent and non-scattering reference samples ($\alpha$ and $\beta$) with known emissivities $\epsilon_{\alpha,\beta}(\lambda)$ and reflectances $R_{\alpha,\beta}(\lambda)$ must be used. More specifically, the total background for $\alpha$ and $\beta$ can be obtained from the measured thermal emission spectra at $T_1$:

$$B_\alpha(\lambda) = \frac{S_\alpha(\lambda, T_1)}{m(\lambda)} - \epsilon_\alpha(\lambda)I_{BB}(\lambda, T_1) \tag{18}$$

$$B_\beta(\lambda) = \frac{S_\beta(\lambda, T_1)}{m(\lambda)} - \epsilon_\beta(\lambda)I_{BB}(\lambda, T_1) \tag{19}$$

Note here that because the optical properties of both references do not change with temperature, $\epsilon_{\alpha,\beta}$ and $B_{\alpha,\beta}$ are temperature independent. Then, $B_1(\lambda)$ and $B_2(\lambda)$ can be found using Eq. 17:

$$B_2(\lambda) = \frac{B_\alpha(\lambda) - B_\beta(\lambda)}{R_\alpha(\lambda) - R_\beta(\lambda)} \tag{20}$$

$$B_1(\lambda) = B_\alpha(\lambda) - R_\alpha(\lambda)B_2(\lambda) \tag{21}$$

We used polished fused-silica and sapphire wafers as references to demonstrate this process, because both references are opaque and have negligible scattering in the spectral range of 5 to 17 μm. Using Eq. 16 and the thermal-emission signal in Fig. 4(a), we obtained the response function $m(\lambda)$ of our FTS system, plotted in Fig. 6(a). The $m(\lambda)$ spectra measured using sapphire and fused silica are identical, as expected. The measured total backgrounds (Eqs. 18-19) are plotted in Fig. 6(b), where both backgrounds are negative. The concept of negative background was discussed in Sec. III and is due to contributions from components after the interferometer of our FTS [18]. Furthermore, the total background is different for sapphire and fused silica, indicating that $B_2(\lambda) \neq 0$.

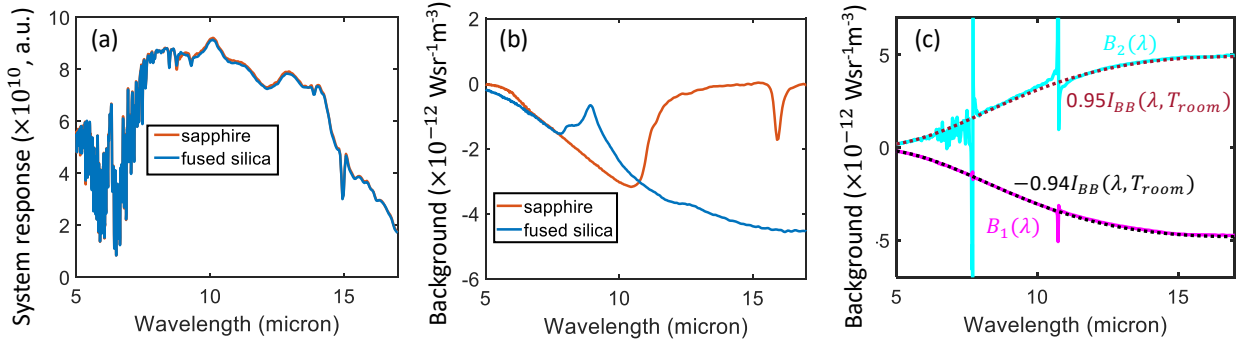

**Figure 6**. (a-b) Measured system response $m(\lambda)$ and background $B(\lambda)$ for sapphire (red) and fused silica (blue) using the thermal-emission spectra shown in Fig. 4(a). (c) Characterized $B_1(\lambda)$ (solid purple) and $B_2(\lambda)$ (solid cyan) of our FTS. $B_2$ and $-B_1$ can be well fitted with theoretical curves for room-temperature emitters with different wavelength-independent effective emissivities (0.94 for $B_1$ and 0.95 for $B_2$).

In Fig. 6(c), we plotted $B_1(\lambda)$ and $B_2(\lambda)$ obtained using Eqs. 20-21. As shown here, $B_2(\lambda)$ is positive while $B_1(\lambda)$ is negative. The positive value of $B_2(\lambda)$ is expected since the background reflected by the emitter into the FTS shares the same path with sample emission: both are detected via transmission through the interferometer (Fig. 7). The negative value indicates that $B_1(\lambda)$ primarily originates from optical components after the interferometer [18] (Fig. 7). This background is detected via reflection by the interferometer without ever reaching the sample; therefore, it is sample-independent. Note that the sharp features in the extracted $B_1(\lambda)$ and $B_2(\lambda)$ near 8 and 11 μm are numerical errors coming from the intersections of the sapphire and fused-silica reflectances: near these two wavelengths, $R_\alpha(\lambda) \sim R_\beta(\lambda)$, resulting in a near-zero denominator in Eq. 20.

The measured $-B_1(\lambda)$ and $B_2(\lambda)$ can be well fitted by the blackbody spectrum from room-temperature emitters with different effective emissivities (0.94 for $B_1$ and 0.95 for $B_2$), as shown in Fig. 6(c). The effective emissivity of $B_2$ (0.95) is expected because the sample is placed inside a sample compartment in our FTS (Fig. 7). The enclosure of the sample compartment can be approximated to be a blackbody with constant emissivity close to unity (Fig. 7) [12]. The effective emissivity of $B_1$ (0.94) indicates that there is a significant amount of background after the interferometer in our FTS (from the "Optical components" in Fig. 7). In principle, $B_1$ can be minimized if the detector is placed immediately after the interferometer. In practice, many FTSs, including our own, are multifunctional commercial systems designed for a variety of measurement modalities. Therefore, many components inside such systems cannot be easily removed, leading to a non-negligible background. Note that $B_1(\lambda)$ and $B_2(\lambda)$ are the total backgrounds that include contributions from various individual components and could change for different setups[2].

---

[2] For example, our setup includes a prism used to couple an external pump laser into the beam path, and we were able to reduce the effective emissivity significantly by removing this component.

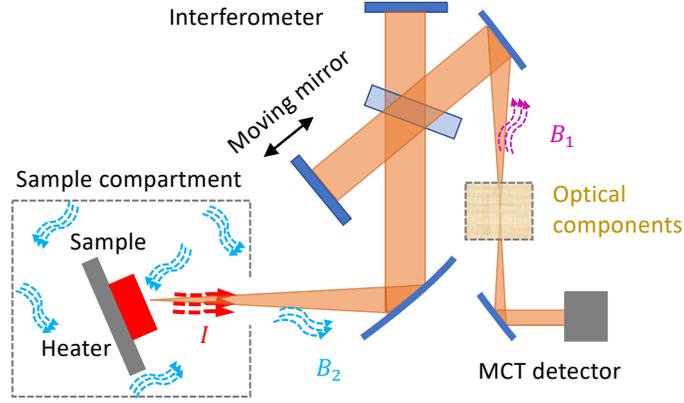

**Figure 7**. Illustration of the various sources of thermal emission, including the thermal emission from the sample ($I$, red arrows), the sample-dependent background ($B_2$, blue arrows), and the sample-independent background ($B_1$, purple arrows). In our FTS, there is only one aluminum mirror (with a very small emissivity) between the sample compartment and the interferometer. Therefore, most of the background from the sample side originates from the sample compartment walls and is reflected by the sample into the FTS beam path ($B_2$). $B_1$ originates from optical components between the interferometer and the detector and is reflected by the interferometer to the detector.

### 1.1: Equilibrium emitters

For an opaque and non-scattering emitter, its emissivity and reflectivity are related via Kirchhoff's law (Eq. 11). Therefore, the measured thermal-emission signal from such an emitter can be further written as

$$S_x(\lambda, T) = m(\lambda)\{\epsilon_x(\lambda, T)I_{BB}(\lambda, T) + [1 - \epsilon_x(\lambda, T)]B_2(\lambda) + B_1(\lambda)\}. \tag{22}$$

With known $m(\lambda)$, $B_1(\lambda)$ and $B_2(\lambda)$, $\epsilon_x(\lambda, T)$ can be determined from $S_x(\lambda, T)$ using Eq. 22:

$$\epsilon_x(\lambda, T) = \frac{\frac{S_x(\lambda,T)}{m(\lambda)} - B_1(\lambda) - B_2(\lambda)}{I_{BB}(\lambda,T) - B_2(\lambda)}. \tag{23}$$

We denote Eq. 23 *Method* 3 for characterizing emissivity for opaque and non-scattering emitters.

Similar expressions have been derived previously [33]–[35] and successfully used for spectrometers on board the Mars exploration rovers [51]. In these works, a blackbody reference at two temperatures was used to find $m(\lambda)$ and $B_1(\lambda)$, while $B_2(\lambda)$ was engineered to be the blackbody distribution at a fixed temperature $T_0$ by engineering the sample compartment to be an approximate blackbody [*i.e.*, $B_2(\lambda) \approx I_{BB}(\lambda, T_0)$] [33], [34]. In ref. [34], a gold diffuse reflector was used to explictly measure $B_2(\lambda)$. In order for this method to work, both the blackbody reference and the sample compartment need to be close to ideal blackbodies (*i.e.*, emissivity very close to 1). Instead, in this paper, we demonstrate equivalent calibration using two arbitrary but known references. Our calibration method also has no requirement on the sample-compartment emissivity, and this should work for any measurement system.

The calibration of our system (Fig. 6) was performed using polished fused-silica and sapphire wafers at 323

and 348 K. As an initial confirmation of *Method 3*, we measured these same samples at 363 K, and extracted their emissivities based on Eq. 23. The resulting data are plotted in Fig. 8(a), and are in good agreement with the same results using *Method 2* (Eq. 10), though *Method 3* yields an emissivity spectrum with less noise.

We then validated *Method 3* by applying Eq. 23 to a temperature-dependent emitter. Figure 8(b) shows the measured emissivity using *Method 3* of the same $VO_2$-based emitter that we studied previously in Sec. IV(2). We also performed reflectance measurements at the same temperatures and plotted the indirectly measured emissivity from Kirchhoff's law (*i.e.*, $1 - R$), which are in excellent agreement with *Method* 3. The slight difference between two measurements is due to the different numerical apertures and measurement angles used for the direct emission and indirect reflection measurements (in our FTS, the numerical aperture and measurement angle for the direct emission are 0.05 and 10°, while they are 0.4 and 0° for indirect reflection measurement). Note that $VO_2$ has a non-negligible hysteresis, which was not observed here because all measurements were performed while increasing the sample temperature.

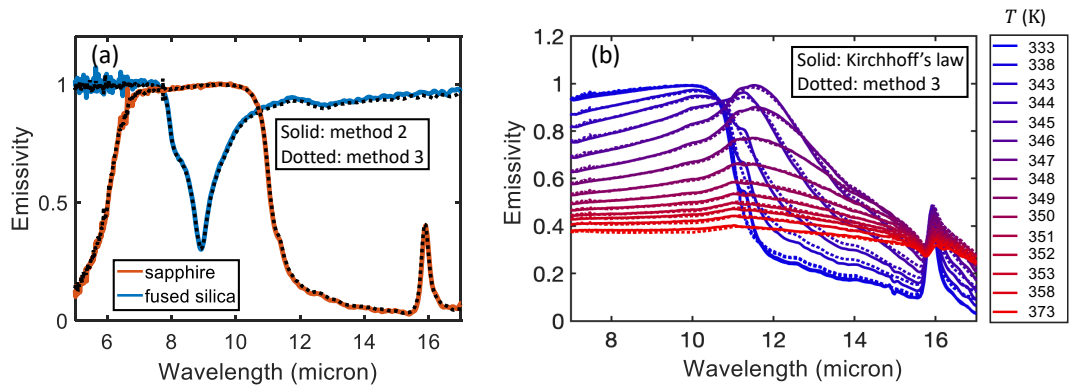

**Figure 8**. (a) Comparison of the measured emissivity using two different methods (solid for *Method 2* and dotted for *Method 3*) of polished wafers of sapphire (red) and fused silica (blue). (b) Comparison of emissivity of the $VO_2$-based emitter from 333 to 373 K, obtained via Kirchhoff's law ($1 - R$, solid) and direct-emission measurements using *Method 3* (dotted). Both direct and indirect measurements were performed for increasing temperature.

### 1.2: Non-equilibrium emitters

The results shown in Fig. 8 are emissivity measurement of samples in thermal equilibrium (*i.e.*, all parts of the sample are at the same temperature). When an emitter is not in thermal equilibrium, a single temperature and emissivity cannot be defined. For example, the temperature of the carriers in a semiconductor or metal can be different from the lattice temperature when pumped by an optical pulse [15], [52]. Also, a macroscopic emitter may have a position-dependent temperature distribution throughout its volume [8]. In such cases, one must measure the true emission signal from the emitter using *Method 3* because the emissivity cannot be defined and thus *Methods 1* and *2* are no longer applicable. In this case, the total thermal emission can be obtained from:

$$I_x(\lambda, T) = \frac{S_x(\lambda,T)}{m(\lambda)} - B_1(\lambda) - R_x(\lambda,T)B_2(\lambda). \tag{24}$$

Therefore, in order to measure the thermal emission from an opaque and non-scattering thermal emitter with a fully calibrated system [*i.e.*, known $m(\lambda)$, $B_1(\lambda)$, and $B_2(\lambda)$], one just needs to measure the thermal emission from the emitter $S_x(\lambda, T)$ and its reflectance $R_x(\lambda, T)$.

Here, we demonstrate a thermal-emission measurement using a flat, homogeneous semitransparent emitter that has a temperature gradient as a function of depth and can therefore be considered a non-equilibrium emitter. We recently demonstrated that the depth-dependent temperature can be extracted from measuring thermal-emission spectra in such a system [53].

We used the natural temperature gradient that exists when samples are heated asymmetrically from one side. Figure 9(a) is a schematic of the measurement that we first reported in Ref. [53], where a 1-mm-thick fused-silica window was heated by a surface heater from the bottom to relative high temperatures of 473 and 573 K. The magnitude of the temperature gradient that is formed is related to the thickness and thermal conductance of the sample, as well as the heater temperature (more discussions in Appendix II). The temperature gradient could be safely neglected for sapphire wafer and the VO$_2$-based emitter in Figs. 3-5 because of the high thermal conductivity of sapphire. The temperature gradient can also be safely neglected for the fused-silica wafer in Fig. 4 because the heater temperature was not very high (< 350 K). The agreement between the calibrated emissivity from the direct-emission measurement and the indirect measurement in Fig. 4 is partially confirms the negligible temperature gradient in the sample.

For heater temperatures of 473 and 573 K, the relatively low thermal conductivity of fused silica ($k \sim 1.4$ W/m·K) created a considerable temperature gradient between the bottom and the top surface, resulting in a non-equilibrium thermal state for the fused-silica window [Fig. 9(a)]. The corresponding thermal emission was measured from the top surface using our FTS and plotted in Fig. 9(b). In the measured data, the desired thermal-emission signal is inevitably mixed with the background. Using the calibration procedure in Eqs. 16-21 and the calibrated system parameters in Fig. 6 and the measured reflectance of the sample, we obtained the true emission spectra using Eq. 14, which are plotted in Fig. 9(c) (reproduced from Ref. [53]).

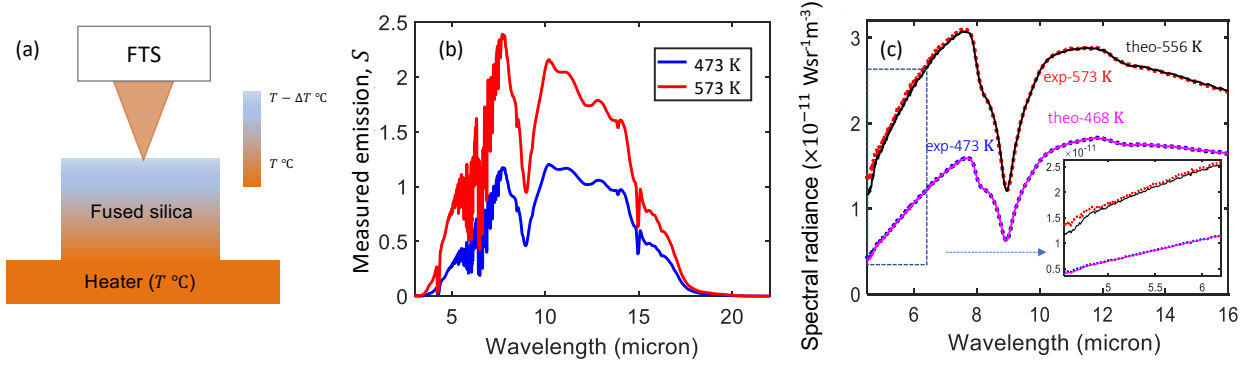

**Figure 9**. (a) Simplified schematic of the experiment for measuring thermal emission from a gradient-temperature (non-equilibrium) emitter. A 1-mm-thick fused-silica wafer was heated by a heater from the bottom. A temperature gradient was formed across the depth of the window due to the low thermal conductivity of fused silica. The corresponding thermal emission was measured by an FTS. (b) Measured thermal-emission signal from the fused-silica window with heater temperature of 473 (blue) and 573 K (red). (c) Characterized thermal emission spectra (dotted) at 473 and 573 K can be well fitted by theoretical emission at 468 and 556 K (solid) for $\lambda > 8$ μm. Panels (b, c) are reproduced from Ref. [53]. Experimental spectra are larger than theoretical values [inset of (c)] for $\lambda < 8$ μm, indicating thermal emission from hotter regions beneath the surface.

We fitted the experimental thermal emission with theoretical predictions which is the product of fused-silica emissivity and the blackbody spectrum. As shown in Fig. 9(c), the theoretically fitted spectra of the fused-silica window overlap almost perfectly with the experimental measurements only for $\lambda > 8$ μm, where fused silica is highly opaque. For $\lambda < 8$ μm, the measured spectra are slightly larger than the theoretical predictions [inset, Fig. 9(c)], indicating the contributions from the hotter regions beneath the surface (more discussion can be found in Ref. [53]).

**(2) Characterization of semi-transparent, non-scattering emitters**

Here, we generalize *Method 3* to non-scattering emitters that can reflect and partially transmit light (*i.e.,* the sample-dependent background includes contributions that are both reflected and transmitted through the sample). In this case, the background can be written as:

$$B_x(\lambda, T) = B_1(\lambda) + R_x(\lambda, T)B_2(\lambda) + Tr_x(\lambda, T)B_3(\lambda). \tag{25}$$

Compared to Eq. 17, we have an additional component that is proportional to the transmittance of the emitter: $Tr_x(\lambda, T)B_3(\lambda)$. In addition to the two non-scattering and opaque references $\alpha$ and $\beta$ to find $m(\lambda)$, $B_1(\lambda)$ and $B_2(\lambda)$, a third non-scattering reference $\gamma$ with non-zero transmittance is needed to determine $B_3(\lambda)$. $B_3(\lambda)$ can be found from $S_\gamma(\lambda, T_1)$ as:

$$B_3(\lambda) = \frac{1}{Tr_x(\lambda)}\left[\frac{S_\gamma(\lambda, T_1)}{m(\lambda)} - \epsilon_\gamma(\lambda)I_{BB}(\lambda, T_1) - R_\gamma(\lambda)B_2(\lambda) - B_1(\lambda)\right]. \tag{26}$$

Once the measurement system is fully characterized with known $m(\lambda)$ and $B_i(\lambda)$, $i = 1,2,3$, $\epsilon_x(\lambda, T)$ can be determined from $S_x(\lambda, T)$ using:

$$\epsilon_x(\lambda, T) = \frac{1}{I_{BB}(\lambda,T)}\left[\frac{S_x(\lambda,T)}{m(\lambda)} - R_x(\lambda,T)B_2(\lambda) - Tr_x(\lambda,T)B_3(\lambda) - B_1(\lambda)\right]. \quad (27)$$

For the case of non-equilibrium emitters, its emission can be obtained accordingly as:

$$I_x(\lambda, T) = \frac{S_x(\lambda,T)}{m(\lambda)} - R_x(\lambda,T)B_2(\lambda) - Tr_x(\lambda,T)B_3(\lambda) - B_1(\lambda). \quad (28)$$

Equation 27 can be further simplified using the fact that $\epsilon_x(\lambda, T)$, $R_x(\lambda, T)$ and $Tr_x(\lambda, T)$ are not independent, rather they are related via Kirchhoff's law for non-scattering objects

$$Tr_x(\lambda, T) + R_x(\lambda, T) + \epsilon_x(\lambda, T) = 1. \quad (29)$$

Substituting Eq. 29 into Eq. 27 further leads to:

$$\epsilon_x(\lambda, T) = \frac{\frac{S_x(\lambda,T)}{m(\lambda)} - R_x(\lambda,T)B_2(\lambda) - [1-R_x(\lambda,T)]B_3(\lambda) - B_1(\lambda)}{I_{BB}(\lambda,T) - B_3(\lambda)}. \quad (30)$$

Therefore, to find $\epsilon_x(\lambda, T)$ we need to measure $R_x(\lambda, T)$ and $S_x(\lambda, T)$. Equation 30 is a slightly generalized form of *Method 3* that can be used for characterizing emissivity for semi-transparent, non-scattering emitters.

### (3) Characterization of scattering emitters

The situation becomes more complicated when the emitter is laterally inhomogeneous and is therefore scattering. For a scattering emitter $x$, the sample-dependence of the background can be also due to light scattered by the sample into the beam path, and can be written as:

$$B_x(\lambda, T) = B_1(\lambda) + R_x(\lambda, T)B_2(\lambda) + Tr_x(\lambda, T)B_3(\lambda) + \iint Sc_x(\lambda, T, \theta, \phi)B_4(\lambda, \theta, \phi)\cos\theta d\theta d\phi. \quad (31)$$

Compared to Eq. 25, here we have an additional term that is related to the scattering of the emitter: $\iint Sc_x(\lambda, T, \theta, \phi)B_4(\lambda, \theta, \phi)\cos\theta d\theta d\phi$. Here $\theta$ and $\phi$ are the azimuth and polar angles representing the relative direction of the background $B_4(\lambda, \theta, \phi)$ respect to the sample (the coordinate system is chosen such that $\theta = \phi = 0$ corresponds to the direction normal to sample surface) and $\cos(\theta)$ accounts for the apparent reduction of area for non-normal directions. $Sc_x(\lambda, T, \theta, \phi)$ (in unit of deg$^{-2}$) is the scattering coefficient describing the portion of light incident on the sample from different angles ($\theta$, $\phi$) that is scattered into the measurement beam path. The integration is performed over all angles. Note that, in principle, we can group the reflection and transmission terms into the scattering term in Eq. 31. Here we prefer to keep them separated so that the parallel to Eq. 25 is more clear, and because we anticipate that many scattering samples will still have a strong specular component.

In the most general case where the background is not isotropic, it is very hard to fully calibrate $B_4(\lambda, \theta, \phi)$ using a limited number of references. In this case, the best practice for measuring scattering emitters is to operate at sufficiently high temperatures that the background is insignificant (*Method 1*) or to set the walls of the sample compartment to approximately the same temperature of the detector to achieve an equilibrium condition between them such that the background can be neglected. This means either using a room-temperature detector (*e.g.*, a DTGS detector) [7], [26] or cooling the sample compartment to match the temperature of the detector.

If the background is roughly isotropic (which can presumably be achieved by appropriately engineering the chamber where the sample is located), *i.e.*, $B_4(\lambda, \theta, \phi) \approx B_4(\lambda) \approx B_3(\lambda) \approx B_2(\lambda)$, it is possible to fully calibrate the system in a way similar to non-scattering emitters. Such calibration of scattering thermal emitters has been discussed before [54], [55]. Under this assumption, Eq. 31 can be rewritten as:

$$B_x(\lambda, T) = B_1(\lambda) + R_x(\lambda, T)B_2(\lambda) + Tr_x(\lambda, T)B_2(\lambda) + B_2(\lambda) \iint Sc_x(\lambda, T, \theta, \phi)\cos\theta d\theta d\phi. \quad (32)$$

Therefore, one can follow the procedure in Eqs. 16-21 to find two unknowns $B_1(\lambda)$ and $B_2(\lambda)$ using two known opaque and non-scattering samples, and $\epsilon_x(\lambda, T)$ can be determined from $S_x(\lambda, T)$ using:

$$\epsilon_x(\lambda, T) = \frac{1}{I_{BB}(\lambda, T)}[\frac{S_x(\lambda, T)}{m(\lambda)} - R_x(\lambda, T)B_2(\lambda) - Tr_x(\lambda, T)B_2(\lambda)$$

$$-B_2(\lambda) \iint Sc_x(\lambda, T, \theta, \phi)\cos\theta d\theta d\phi - B_1(\lambda)]. \quad (33)$$

Equation 33 can be simplified using Kirchhoff's law:

$$Tr_x(\lambda, T) + R_x(\lambda, T) + \epsilon_x(\lambda, T) + \iint Sc_x(\lambda, T, \theta, \phi)\cos\theta d\theta d\phi = 1. \quad (34)$$

Substituting Eq. 34 into Eq. 33 gives:

$$\epsilon_x(\lambda, T) = \frac{\frac{S_x(\lambda, T)}{m(\lambda)} - B_2(\lambda) - B_1(\lambda)}{I_{BB}(\lambda, T) + B_2(\lambda)}. \quad (35)$$

Equation 35 has the same form as Eq. 23, used for extracting emissivity for opaque and non-scattering emitters, which is expected considering the isotropic background condition and only one sample-dependent background term. Note that Eq. 35 can be applied to arbitrary thermal emitters without any prior knowledge of their optical properties. Therefore, it seems that the best practice to for measurement of emissivity from arbitrary samples is to place emitters inside a chamber deliberately engineered to have an isotropic background (e.g., surround the sample with walls that are scattering and at least partially absorbing [33], [34]).

For the case of non-equilibrium thermal emitters, its total thermal emission can be obtained as:

$$I_x(\lambda, T) = \frac{S_x(\lambda,T)}{m(\lambda)} - B_2(\lambda)[R_x(\lambda, T) + Tr_x(\lambda, T) + \iint Sc_x(\lambda, T, \theta, \phi)\cos\theta d\theta d\phi] - B_1(\lambda). \quad (36)$$

This situation is more complicated than measuring emissivity of a scattering emitter (Eq. 35): in addition to system calibration to find $m(\lambda)$, $B_1(\lambda)$, and $B_2(\lambda)$, characterization of the emitter optical properties $R_x$, $Tr_x$, and $Sc_x$ are needed. Among these measurements, the characterization of the scattering coefficient $Sc_x(\lambda, T, \theta, \phi)$ is the most difficult one (see Appendix III).

Before concluding this section, we further comment on the process of calibrating the measurement system. As demonstrated in Sec. V(1), we used three measurements $[S_\alpha(\lambda, T_1), S_\alpha(\lambda, T_2),$ and $S_\beta(\lambda, T_1)]$ to find three unknowns $[m(\lambda), B_1(\lambda)$ and $B_2(\lambda)]$. Very precise measurements of the reference samples are needed to obtain an accurate system calibration because the errors in these three measurements will yield errors in the calibrated system parameters. In practice, it is more favorable to perform more than three reference measurements (*e.g.*, measure thermal emission of references $\alpha$ and $\beta$ at multiple temperatures and/or angles, use more than two references, etc.) so that the calibrated system is more robust against measurement errors.

## VII: Conclusions

This paper is intended to serve as a set of guidelines for performing accurate measurements of the spectrum of far-field thermal emission from surfaces, which in some instances simplifies to measuring the emissivity spectrum. In some cases, the easiest way to measure emissivity is by performing reflectance and transmittance experiments and applying Kirchhoff's law. However, there exist a variety of situations in which Kirchhoff's law cannot be easily applied or does not apply at all, and direct measurements of thermal emission are desirable or required. Other times, it is beneficial to perform both direct thermal-emission measurements and emissivity measurements based on Kirchhoff's law as a check on the accuracy of both approaches.

We found that depending on both the optical properties and temperature-dependence of the emitter and whether the background emission in the instrument can be neglected, different methods should be used for direct measurements of thermal emission. We described several such methods of thermal-emission measurement that address situations of increasing complexity and, in most cases, demonstrated the methods using the Fourier-transform spectrometer (FTS) in our laboratory. In particular, we found that a widely used approach to separating the thermal background and desired signal from thermal-emission measurements fails in instances where the emissivity changes rapidly with temperature—for example, emitters that comprise phase-change materials. Our results show that a measurement apparatus can be fully characterized to enable measurement of such temperature-dependent emissivities, and to measure emitters that are not in equilibrium.

## Acknowledgements

This work is a compilation of measurement processes we developed during projects from the Office of Naval Research (N00014-16-1-2556), the Department of Energy (DE-NE0008680), and the National Science Foundation (ECCS-1750341). The VO$_2$ samples were grown by Zhen Zhang in Shriram Ramanathan's group at Purdue University. We acknowledge helpful discussions with Jonathan King and William Derdeyn.

## Appendix I

Here we show that *Method 2* cannot be applied to emitters that emissivities change with temperature with a rate faster than a certain value. We start with Eq. 12 from Sec. IV., which assumes that the emissivity of the sample stays approximately the same within $\Delta T$ [*i.e.*, $\epsilon_x(\lambda, T + \Delta T/2) \approx \epsilon_x(\lambda, T - \Delta T/2)$]:

$$\epsilon_x(\lambda, T) \stackrel{?}{=} \epsilon_\alpha(\lambda) \frac{S_x(\lambda, T+\Delta T/2) - S_x(\lambda, T-\Delta T/2)}{S_\alpha(\lambda, T+\Delta T/2) - S_\alpha(\lambda, T-\Delta T/2)}. \tag{A1}$$

We expand each term for sample $x$ and reference $\alpha$ in Eq. A1, using Eq. 3

$$S_x(\lambda, T \pm \Delta T/2) = m(\lambda)\{\epsilon_x(\lambda, T \pm \Delta T/2) I_{BB}(\lambda, T \pm \Delta T/2) + B_x(\lambda, T \pm \Delta T/2)\} \tag{A2}$$

$$S_\alpha(\lambda, T \pm \Delta T/2) = m(\lambda)\{\epsilon_\alpha(\lambda) I_{BB}(\lambda, T \pm \Delta T/2) + B_\alpha(\lambda)\} \tag{A3}$$

Note here, the reference $\alpha$ is assumed to have a temperature-independent emissivity and background. To further proceed, we make a first-order Taylor approximation of the temperature-dependent variables:

$$\epsilon_x\left(\lambda, T \pm \frac{\Delta T}{2}\right) \approx \epsilon_x(\lambda, T) \pm \frac{\partial \epsilon_x(\lambda, T)}{\partial T} \frac{\Delta T}{2} \tag{A4}$$

$$I_{BB}\left(\lambda, T \pm \frac{\Delta T}{2}\right) \approx I_{BB}(\lambda, T) \pm \frac{\partial I_{BB}(\lambda, T)}{\partial T} \frac{\Delta T}{2} \tag{A5}$$

$$B_x\left(\lambda, T \pm \frac{\Delta T}{2}\right) \approx B_x(\lambda, T) \pm \frac{\partial B_x(\lambda, T)}{\partial T} \frac{\Delta T}{2} \tag{A6}$$

Substituting all the expansions into Eq. A1 and after some algebra, we obtain the following result:

$$\epsilon_{x\prime}(\lambda, T) = \epsilon_x(\lambda, T) + \left\{I_{BB}(\lambda, T) \frac{\partial \epsilon_x(\lambda, T)}{\partial T} + \frac{\partial B_x(\lambda, T)}{\partial T}\right\} / \frac{\partial I_{BB}(\lambda, T)}{\partial T}. \tag{A7}$$

Equation A7 shows that the measured emissivity $\epsilon_{x\prime}(\lambda, T)$ using *Method 2* (Eq. 12) does not return the true value of $\epsilon_x(\lambda, T)$. Instead, there is an additional term related to the temperature-dependent emissivity and background and the temperature of measurement.

Equation A7 is very general and applies to any measurement setup. Specific to our FTS, Eq. A7 can be further simplified. As discussed in Sec. V(1), the total background $B_x$ for *our* FTS can be rewritten as $-0.94 I_{BB}(\lambda, T_{room}) + 0.95 I_{BB}(\lambda, T_{room}) R_x(\lambda, T)$, where $R_x(\lambda, T) = 1 - \epsilon_x(\lambda, T)$ according to Kirchhoff's law. Therefore, $\partial B_x(\lambda, T)/\partial T = -0.95 I_{BB}(\lambda, T_{room}) \partial \epsilon_x(\lambda, T)/\partial T$. Accordingly, Eq. A7 can be written as follows for an opaque and non-scattering emitter in our specific FTS:

$$\Delta \epsilon_x(\lambda, T) = \epsilon_{x\prime}(\lambda, T) - \epsilon_x(\lambda, T) = \{I_{BB}(\lambda, T) - 0.95 I_{BB}(\lambda, T_{room})\} \frac{\partial \epsilon_x(\lambda, T)}{\partial T} / \frac{\partial I_{BB}(\lambda, T)}{\partial T}. \tag{A8}$$

In Eq. A8, only $\epsilon_x$ is related to the sample, and the remaining parts are all known. Therefore, Eq. A8 can be further written as:

$$\Delta \epsilon_x(\lambda, T) = H(\lambda, T) \frac{\partial \epsilon_x(\lambda, T)}{\partial T}, \tag{A9}$$

where $H(\lambda, T) = \{I_{BB}(\lambda, T) - 0.95 I_{BB}(\lambda, T_{room})\}/ \partial B_x(\lambda, T)/\partial T$. We plot $H(\lambda, T)$ for our FTS in the Fig. A1. As shown here, for sample temperatures between 343 and 373 K, $H(\lambda, T)$ is positive and does not change significantly with wavelength. Figure A1 also tells that with the same rate of emissivity change, a higher sample temperature will lead to a larger error using *Method 2*.

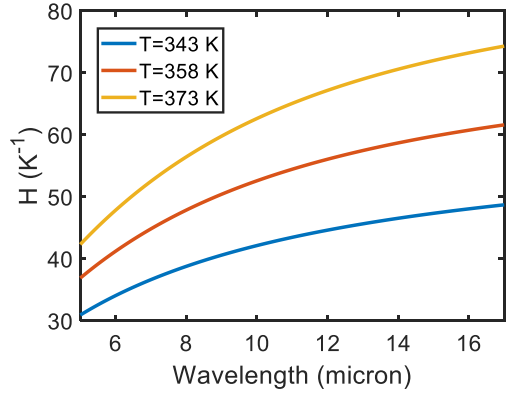

**Figure A1**. $H(\lambda, T)$ as a function of wavelength for different sample temperatures for our FTS.

## Appendix II

Here we estimate the magnitude of temperature gradient for different samples such as those in Figs. 3 and 4 when they are heated from the bottom [Fig. 9(a)]. The temperature distribution can be obtained by finding the steady solution of the following one-dimensional heat transfer equation [15]:

$$\frac{\partial T(t,z)}{\partial t} = \frac{kA}{\rho C_p} \frac{\partial^2 T(t,z)}{\partial z^2} \tag{A10}$$

where $k, \rho, C_p$ and $A$ represent the thermal conductivity, density, heat capacity, and area of the sample, respectively. $z$ is the spatial coordinate along the vertical direction. The boundary conditions used to solve Eq. A10 are (1) a conductive boundary at the bottom (*i.e.*, sample temperature at the bottom is the same as the temperature of the heater) and (2) a convective boundary with the surrounding room-temperature air (*i.e.*, the heat-transfer rate is $hA\Delta T$, where $h$ is the heat-transfer coefficient of the sample–air boundary and $\Delta T$ is temperature difference between the top surface of the sample and the ambient air). Note that the exact value of $h$ is hard to determine because it changes with the sample geometry and it also increases with the sample temperature (*i.e.*, a hotter sample will lead to a stronger air flow, resulting a larger $h$) [56]. Here we do not explicitly consider the radiative heat transfer at the top surface, and its impact is grouped into the value of $h$. Note that we are not trying to reproduce our experimental measurement here because our understanding of the convection boundary condition is very limited, and the analysis here only provides an order-of-magnitude estimate.

Figure A2(a) shows the calculated steady state temperature distribution for 1-mm-thick fused silica and sapphire slabs for heater temperature of 373 K. In the calculation, $k = 1.4$ W/m·K, $\rho = 2.2$ g/cm³, $C_p = 740$ J/Kg·K are chosen for fused silica [57] and $k = 24$ W/m·K, $\rho = 4.0$ g/cm³, $C_p = 760$ J/Kg·K are chosen for sapphire [58]. A reasonable value of $h = 10$ W/m·K is chosen for the convection boundary, assuming still air [56]. As shown here, the temperature drop across the entire 1-mm slab is only about 0.5 K for fused silica and is negligibly small for sapphire. Figure A2(b) shows the calculated steady state temperature when both samples are heated to 573 K. Here, $h = 80$ W/m·K is chosen such that a temperature drop about 17 K is obtained for the fused-silica wafer [Fig. A2(b)], which agrees with our experimental measurement [53]. Note that this value of $h$ is an effective convection coefficient that here includes radiative heat transfer from the surface. Even with such a large value of $h$, the temperature drop across the entire sapphire slab is still less than 1 K due to its much higher thermal conductivity.

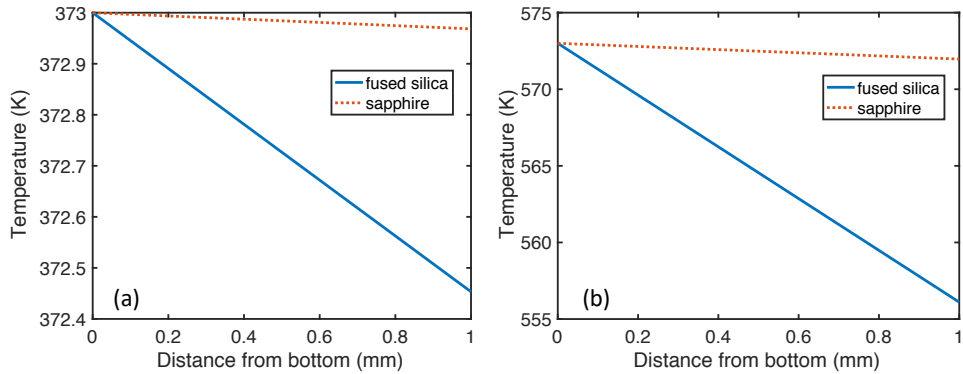

**Figure A2**. Calculated temperature distribution for 1-mm thick fused-silica (solid blue) and sapphire slabs (dotted red) when they are heated from the bottom by a heater with temperatures of (a) 373 and (b) 573 K, surrounded by ambient air.

## Appendix III

To measure non-equilibrium scattering emitters using *Method 3*, the scattering coefficient of the sample

$Sc_x(\lambda, T, \theta, \phi)$ must be independently measured. This can be done with a setup such as the one in Fig. A3, sometimes referred to as an angle-resolved scatterometer [59].

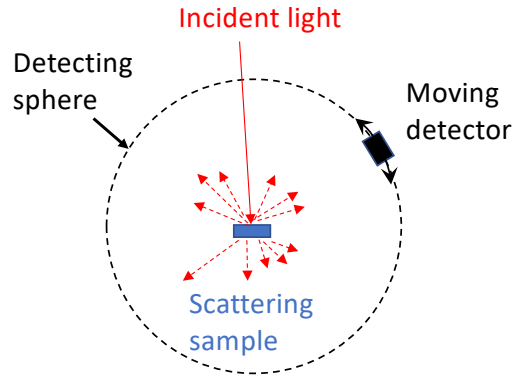

**Figure A3**. Schematic of an angle-resolved scatterometer. Light is incident from a certain direction and a detector moves along a detecting sphere to measure the scattered light over all angles $(\theta, \phi)$. Because of reciprocity, the scattering coefficient of the sample, $Sc_x(\lambda, T, \theta, \phi)$, can be determined from this measurement.